\documentclass[pre,groupedaddress,showkeys,showpacs,twocolumn]{revtex4}

\usepackage{amsmath}
\usepackage{graphics}
\usepackage{graphicx}
\usepackage{amsfonts}
\usepackage{amssymb}
\usepackage{dcolumn}
\usepackage{bm}
\usepackage{epstopdf}
\begin{document}
\title{Contact of a spherical probe with a stretched rubber substrate}
\author{Christian Fr\'etigny}
\author{Antoine Chateauminois}
\email [] {antoine.chateauminois@espci.fr}
\affiliation{Soft Matter Science and Engineering Laboratory (SIMM),PSL Research
University, UPMC Univ Paris 06, Sorbonne Universités, ESPCI Paris, CNRS, 10 rue Vauquelin,
75231 Paris cedex 05, France}
\date{\today}
\begin{abstract}
We report on a theoretical and experimental investigation of the normal contact of stretched neo-Hookean substrates with rigid spherical probes. Starting from a published formulation of surface Green's function for incremental displacements on a pre-stretched, neo-Hookean, substrate (L.H. Lee \textit{J. Mech. Phys. Sol.} \textbf{56} (2008) 2957-2971), a model is derived for both adhesive and non-adhesive contacts. The shape of the elliptical contact area together with the contact load and the contact stiffness are predicted as a function of the in-plane stretch ratios $\lambda_x$ and $\lambda_y$ of the substrate. The validity of this model is assessed by contact experiments carried out using an uniaxally stretched silicone rubber. for stretch ratio below about 1.25, a good agreement is observed between theory and experiments. Above this threshold, some deviations from the theoretical prediction are induced as a result of the departure of the mechanical response of the silicone rubber from the neo-Hokeean description embedded in the model.
\end{abstract}
\maketitle
%
\section*{Introduction}
\label{sec:Introduction}
Contact of soft solids such as elastomers, gels or biological tissues with rigid probes pertains to many practical situations including, as an example, the determination of the mechanical properties of these objects using indentation methods. Such systems being very easily deformed, contact stresses are often superimposed to bulk stresses which fall beyond the limit of linear elastic descriptions. As an example, one can cite the local friction of smooth rubbers surfaces with statistically rough rigid bodies. At the scale of the macroscopic contact, the finite sizes of the contacting bodies induce in-plane surface strains which can easily exceed 0.2 can under the action of a frictional stress~\cite{nguyen2011,nguyen2013}. At the microscopic scale, this implies that single micro-asperity contacts occur locally on a  pre-stretched rubber surface. The effects of such finite strains on micro-contacts shape and stresses are largely overlooked in current contact mechanics description of rough contacts, although they may affect the prediction of the actual contact area and the associated frictional forces.\\
\indent From a theoretical perspective, contact problems on soft rubber substrates subjected to finite strains have been essentially handled within the framework of the infinitesimal deformation theory developed by Biot~\cite{biot1965}, Green and co-workers~\cite{green1952,green1954}. In these approaches, contact-triggered infinitesimal deformations are superimposed upon finite deformations due to pre-stress. Early solutions along these lines include the work by Dhaliwal~\cite{dhaliwal1980,dhaliwal1978} and co-workers who handled the problem of the contact of rigid axisymmetric probes with a neo-Hookean body under a state of uniform biaxial stretching. Using the solution established by Dhaliwal and Singh~\cite{dhaliwal1978}, Yang derived analytical expressions for the relation between contact stiffness, contact area, elastic constants, and finite stretch~\cite{yang2004}. Here again, the theory deals with the axisymmetrical indentation of a neo-Hoohean solid under uniform bi-axial stretching. Additional solutions for plane-strain contacts with hyperelastic half spaces were also derived by Brock~\cite{brock,brock1999} which incorporate anisotropic frictional situations.\\
\indent These problems were experimentally addressed by Barquins and co-workers who carried out a series of contact experiments involving spherical or cylindrical probes and natural rubber sheets under a state of either uniaxial or uniform biaxial tensile strains~\cite{barquins1976,barquins1993,barquins1991,felder1992}. These works were carried out with the objective of investigating the effects of a pre-stretch on the formation of Schallamach waves~\cite{barquins1976}, on rolling friction~\cite{barquins1993} and on adhesion~\cite{barquins1991,felder1992}. They especially evidenced the anisotropy induced by uniaxial tensile stretching which result in the development of elliptical contact shapes (with spherical probes) and in anistotropic friction forces. These experiments were revisited later on by Gay~\cite{gay2000} which assumed that a superposition principle can be applied to the response of the rubber to both the initial stretching and the deformation due to the rigid probe. This superposition being performed in Lagrangian coordinates, Gay developed an argument stating that the indentation stage of an uniaxially pre-stretched substrate can be assimilated to  the indentation of an elastic half-space by an ellipsoidal indenter which indeed account for the elliptical contact shape.\\
\indent In the present study, we develop a more general approach of the contact problem of a pre-stretched neo-Hookean substrate with a rigid spherical probe. It is based on the Green's function of the solid which describes its response to a point force. Knowledge of this function allows one to calculate the response of the solid to an arbitrary force distribution as the weighted sum of point force responses. In a recent paper, Biggins \textit{et al}~\cite{biggins2014} developed a linear theory with perfect volume conservation for the Green's function which enforce the constraint of isochoric deformations exactly. This approach is found to remain valid until strains become geometrically large but it is only generalizable to 2-D or axisymmetric situations. Here, we make use of the expressions of the Green's function for a neo-Hookean substrate which were recently derived by He~\cite{he2008,he2009} within the framework of incremental strain theory in order to develop a contact model able to handle non axial stretch situations. Solutions are provided for the contact shape, load and contact stiffness which include the effects of adhesion. This model is validated by contact experiments between a spherical glass probe and a silicone substrate under various extent of uni-axial stretching.
\section*{Experimental details}
\label{sec:experimental_details}
\begin{figure}[ht]
	\begin{center}
		\includegraphics[width=0.9\columnwidth]{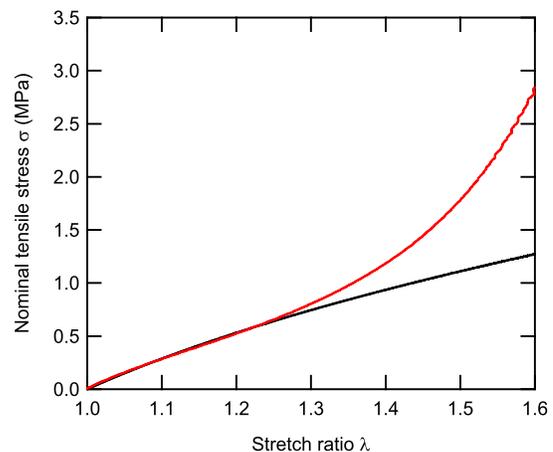}
		\caption{(Color online) Nominal tensile stress as a function of stretch ratio of the PDMS rubber (cross-head speed: 0.5~mm s$^-1$). The black solid line corresponds to a fit of data points for $\lambda <1.2$ to a neo-Hookean model (Eq.~(\ref{eq:neo_hooke})) with $C_1=0.526$~MPa.} 
		\label{fig:stress_stretch}
	\end{center}
\end{figure}

\begin{figure}[h]
	\begin{center}
		\includegraphics[width=0.9\columnwidth]{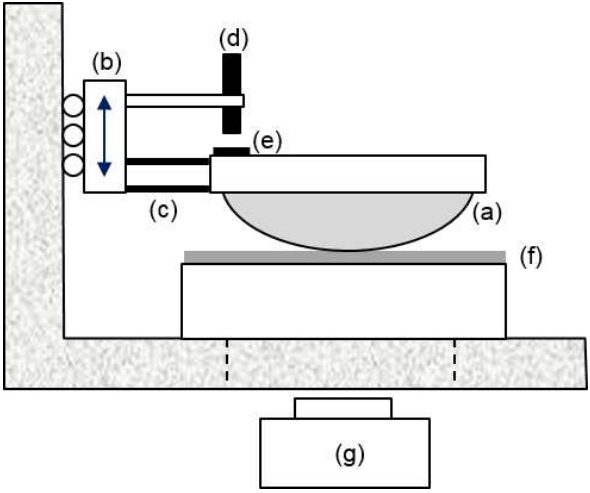}
		\caption{Schematic of the custom-built indentation set-up. A spherical indenter (a) is fixed to a vertical translation stage (b) by means of a cantilever with two flexible arms (c). During the application of normal contact loading, a measurement of the deflection of the cantilever by means of an optical fiber (d) and a  reflecting surface (e) allows to determine the applied normal load on the PDMS substrate (f). Images of the contact region are recorded with a microscope and a camera (g).} 
		\label{fig:setup}
	\end{center}
\end{figure}

A commercially available  transparent Poly(DiMethylSiloxane) silicone (PDMS Sylgard 184, Dow Corning, Midland, MI) is used as an elastomer substrate. The silicon monomer and the hardener are mixed in a 10:1 weight ratio and crosslinked at 70~$^\circ C$ for 48 hours. In order to accurately monitor the level of surface stretching in the contact zone, a square network of small cylindrical holes (diameter 20 $\mu$m, depth 5 $\mu$m and center-to-center spacing 80 $\mu m$) is stamped on the PDMS surface by means of standard soft lithography techniques. Once imaged in transmission with a white light, the pattern appears as a network of dark spots which are easily detected. Full details regarding the design and fabrication of PDMS substrates are provided in ~\cite{nguyen2011}.\\
\indent The stretching behavior of the PDMS rubber was determined using a conventional tensile testing machine (Instron 5565) equipped with an optical extensometer. Dog-bone shaped specimens with a gage length 28x4x2~mm$^3$ were loaded at an imposed cross-head speed of 0.5~mm s$^{-1}$ up to a stretch ratio of 1.6. The resulting nominal stress $\sigma$ versus stretch ratio $\lambda$ response is shown in Fig.~\ref{fig:stress_stretch}. Data up to $\lambda=1.25$ (i.e. before the occurrence of significant strain hardening) have been fitted using a neo-Hookean model
  \begin{equation}
  \sigma=2C_1\left(\lambda - \frac{1}{\lambda^2}\right) \:,
  \label{eq:neo_hooke}
  \end{equation}
with $C_1=0.526$~MPa. The resulting fit is reported as a black line in Fig.~ \ref{fig:stress_stretch}.\\
\indent Contact experiments are carried out using PDMS substrates $5~\times~30~\times~100$~mm$^{3}$ and a plano-convex BK7 glass lens with a radius of curvature of 5.2~mm (Melles Griot, France).  A schematic of the custom set-up is shown in Fig.~\ref{fig:setup}. The lens indenter is fixed to a vertical translation stage (Microcontrole, UMR 8.25) by means of a double cantilever. An optical fiber (Philtec, Model D25) mounted on the vertical stage, allows to measure with sub-micrometer resolution the deflection of the blades during the indentation process. A mirror is located on the cantilever tip which provides a reflecting surface for the displacement sensor. Then, from a knowledge of the calibrated stiffness of the cantilever (11.7~N~m$^{-1}$), the applied normal load can be determined with mN accuracy from the measured deflection of the cantilever. All the experiments are carried out  with normal loads less than 150~mN. These load range ensures the achievement of semi-infinite contact conditions (\textit{i.e.} the ratio of substrate thickness to contact radius is larger than ten \cite{gacoin2006}). In order to vary the extent of adhesive forces between surfaces, some experiments were carried out with the contact fully immersed within a droplet of deionized water.\\
\indent A zoom lens mounted on a CMOS camera (PhotonPhocus, MV1024E) records 1024~x~1024, 8 bits images of the contact region through the transparent PDMS substrate. When stretched, the PDMS substrate is fixed between two grips and the stretch ratio is measured optically from the deformation of the dot pattern at the surface of the rubber specimen. Indentation experiments are carried out using a step-by-step loading procedure. At each load step, a contact image is recorded after ensuring that the contact size is no longer evolving due to adhesive effects.
\section*{Contact model}
 \label{sec:contact_model}
We consider the normal contact between a rigid spherical probe with radius $R$ and a neo-Hookean elastomer substrate (shear modulus $\mu$) which undergoes a uniform, finite pre-stretch $\lambda_x, \lambda_y$ along the two orthogonal directions $x$ and $y$ along the surface plane. In order to establish the contact equations, we take advantage of the work by He~\cite{he2008} who derived a surface Green function for incremental displacements on a pre-stretched incompressible substrate obeying a neo-Hookean constitutive law. Using this approach, the characteristics of the contact are deduced below under the assumptions that the contact is frictionless and that the coupling between normal pressure and lateral displacements - which does not cancel, contrarily to the elastic case -  has negligible effects on the results.

Guided from experimental results, we assume that the contact area is elliptic (semi-major axis $a$ and $b$ along the directions $x$ and $y$ respectively). We further suppose that the normal stress at the surface of the substrate can be derived from a one-dimensional function through scaling along both axis, i.e. a function $p\left( u \right)$ can be defined which cancels for $u>1$ from which the normal stress $\sigma_{zz}$ can be expressed as
\begin{equation}
	\sigma_{zz}\left( \textbf{r} \right)=p\left( \sqrt{\frac{x^2}{a^2}+\frac{y^2}{b^2}} \label{eq:pression}\right) \,.
\end{equation}
\indent It is shown below that a stress distribution in this form may indeed give rise to paraboloidal vertical displacements with cylindrical symmetry on a pre-stretch substrate; it can thus be a solution to the small indentation problem of a sphere. Normal displacements on the pre-stretched substrate can be accounted for by using the appropriate Green tensor coefficient $G_{zz}$, which can be written in the two-dimensional Fourier transform space as~\cite{he2008}
\begin{equation}
	G_{zz}\left( \textbf{k} \right)=\frac{16\lambda_x^2\lambda_y^2}{3K}\frac{\left(  \eta+k\right)k}{\omega} \,,
\end{equation}
where
\begin{align}
	\eta&=\lambda_x\lambda_y\sqrt{\lambda_x^2k_x^2+\lambda_y^2k_y^2} \,,\\
	\omega&=\eta^3+\eta^2k+3\eta k^2-k^3\,,
\end{align}
and $K$ is a reduced modulus often used in contact mechanics~\cite{johnson1985}; for incompressible materials, $K=\frac{16}{3}\mu$ where $\mu$ is the shear modulus. Normal displacements $u_z\left( \textbf{r} \right)$ are deduced from the inverse Fourier transform
\begin{equation}
	u_z\left( \textbf{r} \right)=\frac{1}{4\pi^2}\iint G_{zz}\left( \textbf{k} \right)\hat{\sigma}_{zz}\left(\textbf{k} \right)e^{-i\textbf{k}\textbf{r}}dk_xdk_y \,,
\end{equation} 
where $\hat{\sigma}_{zz}\left(\textbf{k} \right)$ is the Fourier transform of the normal stress, which can be expressed using the similarity property as
\begin{align}
	\hat{\sigma}_{zz}\left( \textbf{k} \right)&=\iint p\left( \sqrt{\frac{x^2}{a^2}+\frac{y^2}{b^2}} \right) e^{i\textbf{k}\textbf{r}}dxdy\\
	&= ab\int_0^1 udu\int_0^{2\pi}  p\left(u\right) e^{ik\gamma u\cos v}\,dv   \\
	&=2\pi ab \int_0^1  p\left(u\right)  J_0\left( k\gamma u \right)u\,du\\
	&=2\pi ab\hat{\sigma}\left( k\gamma \right) \,,
\end{align}
where
\begin{align}
	\gamma&=\sqrt{a^2\bar{k}_x^2+b^2\bar{k}_y^2} \,,\\
	k&=\sqrt{k_x^2+k_y^2} \,.
\end{align}
$J_0\left( . \right)$ is the Bessel function of order 0 and the components of the unit vector $k^{-1}\textbf{k}$  along the axis $x$ and $y$ are noted $\bar{k}_x$ and $\bar{k}_y$ respectively. Then, Fourier transform of the normal stress, $\hat{\sigma}\left(  \textbf{k}\right)$, is expressed from the Hankel transform of the one-dimensional stress function $p\left(u\right)$. The polar angles of $\textbf{k}$ and $\textbf{r}$ are noted respectively $\beta$ and $\theta$. Normal displacements can be expressed as
\begin{align}
	u_z\left( \textbf{r} \right)=ab\frac{8\lambda_x^2\lambda_y^2}{3\pi K}\int_0^{2\pi}\frac{\bar{\eta}+1}{\bar{\omega}}\,d\beta\int_0^\infty \hat{\sigma}\left( k\gamma \right) e^{-ikr\cos\left(\beta-\theta  \right)}\,dk 
\end{align}
where 
\begin{align}
	\bar{\eta}&=\lambda_x\lambda_y\sqrt{\lambda_x^2\cos^2\beta+\lambda_y^2\sin^2\beta} \,,\\
	\bar{\omega}&=\bar{\eta}^3+\bar{\eta}^2+3\bar{\eta}-1 \,.
\end{align}
As the function to be back transformed $G_{zz}\left( \textbf{k} \right)\hat{\sigma}_{zz} \left(\textbf{k} \right)$ is invariant through the change $\textbf{k}\rightarrow -\textbf{k}$, one can drop the imaginary term in the integral and express
\begin{align}
	u_z\left( \textbf{r} \right)=ab\frac{8\lambda_x^2\lambda_y^2}{3\pi K}&\int_0^{2\pi} \frac{\bar{\eta}+1}{\bar{\omega}}\,d\beta  \nonumber\\
	\int_0^\infty & \hat{\sigma}\left( k\gamma \right)\cos{\left(kr\cos\left(\beta-\theta  \right)  \right)}\,dk \label{eq:uzint} 
\end{align}
which includes a cosine transform. As expected from the Fourier-Hankel-Abel (FHA) cycle~\cite{bracewell2000}, expressing stress as an Hankel transform and using Eq.~(6.671.2) in ref.~\cite{gradshteyn1965}, inner integral in the RHS of Eq.~\ref{eq:uzint} can be written as an Abel transform:
\begin{align}
	I&=\int_0^\infty \hat{\sigma}\left( k\gamma \right)\cos{\left(kr\cos\left(\beta-\theta  \right)  \right)}\,dk
	\\
	&=\int_0^1p\left( u \right)udu\int_0^\infty J_0\left( k\gamma u \right)\cos{\left(kr\cos\left(\beta-\theta  \right) \right)}\,dk\\
	&=\frac{1}{\gamma}\int_{\frac{r\left|\cos\left(\beta-\theta  \right)\right|}{\gamma}}^1\frac{up\left( u \right)}{\sqrt{u^2-\frac{r^2\cos^2\left(\beta-\theta  \right)}{\gamma^2}}} \,du
\end{align}
when $r\left|\cos\left(\beta-\theta  \right)\right|\le\gamma$ and $I=0$ when $r\left|\cos\left(\beta-\theta  \right)\right|\ge\gamma$.
Defining the Abel transform of the stress function for $0\le s\le 1$ as
\begin{equation}
	H\left( s \right)=\int_{s}^1\frac{up\left( u \right)}{\sqrt{u^2-s^2}} \,du \label{eq:Abel}
\end{equation}
and $H\left( s \right)=0$ otherwise, the vertical displacement can be expressed as follows
\begin{equation}
	u_z\left( \textbf{r} \right)=ab\frac{8\lambda_x^2\lambda_y^2}{3\pi K}\int_0^{2\pi}\frac{\bar{\eta}+1}{\bar{\omega}\gamma}H\left( \frac{r\left|\cos\left(\beta-\theta  \right)\right|}{\gamma} \right)\,d\beta \,.
	\label{eq:uz}
\end{equation}
The function $H\left(\frac{ r\left|\cos\left( \beta-\theta \right)\right|}{\gamma }\right)$ does not cancel if
\begin{equation}
	r\left|\cos\left( \beta-\theta \right)\right|\le \sqrt{a^2\cos^2\beta+b^2\sin^2\beta} \,.
	\label{eq:condit}
\end{equation}
\indent It can be shown that this condition is fulfilled, whatever the angle $\beta$ is,  for all the points $\textbf{r}$ inside the contact area. Indeed, condition  (\ref{eq:condit}) reads $D\le 0$ where
\begin{equation}
	D=r^2\cos^2\left( \beta-\theta \right)-\left( a^2\cos^2\beta+b^2\sin^2\beta \right) \,.
\end{equation}
For a given point $\left( r,\theta \right)$ situated at the contact edge, the parametric representation of the ellipse implies the existence of an angle $\alpha$ which verifies $r\cos\theta=a\cos\alpha$ and $r\sin\theta=b\sin\alpha$. Then
\begin{equation}
	r^2\cos^2\left( \beta-\theta \right)=\left( a\cos\alpha\cos\beta+b\sin\alpha\sin\beta \right)^2
\end{equation} 
and thus, for this point,
\begin{equation}
	D= -\left( a\sin\alpha\cos\beta-b\cos\alpha\sin\beta \right)^2\le 0 \,.
\end{equation}
For points situated in the contact area, the condition is a fortiori fulfilled.\\
\indent We now define two anisotropy parameters, $\varphi$ and $\psi$ which characterize the stretch state and the eccentricity of the elliptic contact area, respectively:
\begin{align}
	& \varphi=\frac{\lambda_x^2-\lambda_y^2}{\lambda_x^2+\lambda_y^2}; \psi=\frac{{a^2-b^2}}{a^2+b^2} \,,\\
	&\bar{\eta}=\lambda_x\lambda_y\sqrt{\frac{\lambda_x^2+\lambda_y^2}{2}}\sqrt{1+\varphi\cos 2\beta} \,,\\
	&\gamma=c\sqrt{1+\psi\cos 2\beta} \,,\\
	&c=\sqrt{\frac{a^2+b^2}{2}} \,.
\end{align}
\indent In the following, we note $\delta$ the normal displacement of the apex of the spherical indenter below the substrate plane. It will be assumed that the contact size is much smaller than the curvature radius of the indenter and that in-plane displacements can be neglected as compared to normal displacements. In such a situation, normal displacements within the contact area can be assumed to obey an axisymmetrical parabolic dependence to the distance from the apex. Accordingly, they can be expressed as
\begin{equation}
	\delta-\frac{r^2}{2R}=\frac{8\lambda_x^2\lambda_y^2}{3\pi K}\frac{ab}{c}\int_0^{2\pi}\frac{\bar{\eta}+1}{\bar{\omega}\bar{\gamma}}H\left(\frac{r}{c}\frac{ \left|\cos\left( \beta-\theta \right)\right|}{\bar{\gamma} }\right)\,d\beta \label{eq:eqint} \,,
\end{equation}
where $\bar{\gamma}=\sqrt{1+\psi\cos 2\beta}$. This expression constitutes a linear integral equation for the function $H$, of the first kind with constant limits of integration. A polynomial solution can be found~\cite{polyanin2008} in the form
\begin{align}
	& H\left( s \right)=\frac{3K}{8\lambda_x^2\lambda_y^2}\frac{c}{ab}\left( \frac{\delta}{C_0}-\frac{c^2s^2}{2RC_2} \right) \label{eq:H} \,,\\
	& C_m=\frac{1}{\pi}\int_0^{2\pi}\frac{\bar{\eta}+1}{\bar{\omega}\bar{\gamma}}\left(\frac{ \left|\cos\left( \beta-\theta \right)\right|}{\bar{\gamma} }\right)^m\,d\beta \,,
	\label{eq:Cm}
\end{align}
with $m=0,2$. Stress distribution is retrieved using inversion of the Abel transform Eq.~(\ref{eq:Abel}):
\begin{align}
	p\left( u \right)&=-\frac{2}{\pi}\frac{1}{u}\frac{d}{du}\int_u^1
	\frac{sH\left( s \right)}{\sqrt{s^2-u^2}}\,ds \label{eq:contr0}\\
	&=\frac{2}{\pi}\left[\frac{H\left( 1 \right)}{\sqrt{1-u^2}}-\int_u^1\frac{H'\left( s \right)}{\sqrt{s^2-u^2}}\,ds \right] \,, \label{eq:contr}
\end{align}
where $H'\left( s \right)=dH\left( s \right)/ds$. Normal load is obtained by integration of the stress function Eq. (\ref{eq:pression}) expressed by (\ref{eq:contr0}):
\begin{align}
 P&=\iint p\left( \sqrt{\frac{x^2}{a^2}+\frac{y^2}{b^2}} \right)\,dxdy\\
&=2\pi ab\int_0^1 p\left( u \right)u\,du\\
&=4ab\int_0^1 H\left( s \right)\,ds \label{eq:PdeH}
\end{align}
\indent When adhesion is neglected, normal stress is not singular at the contact edge and thus $H\left( 1 \right)=0$. In the Johnson, Kendall and Roberts (JKR) adhesion theory~\cite{johnson1971}, adhesion induces a stress singularity in this region, \textit{i.e.} $H\left( 1 \right)\ne 0$. Both situations are discussed in the following sections. It can be noticed that $H$ function is very similar to the auxiliary function defined by Sneddon to describe the contact of an axisymmetric punch on a flat~\cite{sneddon1965}, which was generalized to the adhesive case~\cite{maugis1981,barquins1982,basire1998}.
\subsection*{Non-adhesive contact}
When $H\left( 1 \right)=0$, it comes from Eqs.~(\ref{eq:H}) and (\ref{eq:contr}) that the normal stress can be derived from Eq. (\ref{eq:pression}) using the function
\begin{equation}
	p\left( u \right)=\frac{3K}{4\pi RC_2\lambda_x^2\lambda_y^2}\frac{c^3}{ab}\sqrt{1-u^2}\label{eq:sigb} \,,
\end{equation}
which exhibits the classical Hertzian shape. To be admissible, the stress field derived from this function must give rise to axi-symmetrical displacements within the contact area. From Eq.~(\ref{eq:H}), it comes that the condition that $C_2$ does not depend on the orientation $\theta$ is sufficient to fulfil this requirement. Expressing 
\begin{equation}
	\cos^2\left( \beta-\theta \right)=
	\frac{1}{2}\left( 1+\cos 2\beta\cos 2\theta \right)+\frac{1}{2}\sin 2\beta\sin 2\theta \,,
\end{equation}
the second right hand term, with a factor $\sin2\beta$, gives a vanishing contribution to the integral~(\ref{eq:Cm}) due to symmetry reasons: the integrand changes sign when $\beta\rightarrow 2\pi-\beta$. One thus obtains
\begin{equation}
	C_2=\frac{1}{2\pi}\int_0^{2\pi}\frac{\bar{\eta}+1}{\bar{\omega}\bar{\gamma}^3}\left( 1+\cos 2\beta\cos 2\theta \right) \,d\beta \,.
\end{equation}
\indent In order to obtain an isotropic result, the contribution of the $\cos2\theta$ term to the integral should vanish, i.e., using the symmetry properties of the integrand,
\begin{equation}
	\int_0^{\frac{\pi}{2}}\frac{\bar{\eta}+1}{\bar{\omega}\bar{\gamma}^3}\cos 2\beta \,d\beta=0 \label{eq:eq} \,.
\end{equation}
For a given stretch state, the functions $\bar{\eta}$ and $\bar{\omega}$ are determined. The solution $\psi$ of this equation, easily obtained using numerical integration, is noted $\psi_s$. For this solution, we define
\begin{align}
	C_{2s}&= \frac{2}{\pi}\int_0^{\frac{\pi}{2}}\frac{\bar{\eta}+1}{\bar{\omega}\bar{\gamma}_s^3} \,d\beta \,,\\
	C_{0s}&=\frac{4}{\pi} \int_0^{\frac{\pi}{2}}\frac{\bar{\eta}+1}{\bar{\omega}\bar{\gamma}_s} \,d\beta \,,\\
	\bar{\gamma}_s&=\sqrt{1+\psi_s\cos 2\beta} \,.
\end{align}
From the condition (\ref{eq:eq}) and the definition of $\bar{\gamma}$, it can be verified that $C_{0s}=2C_{2s}$. Furthermore, exchanging the stretches $\lambda_x$ and $\lambda_y$ gives a solution $\psi_s$ which is opposite (the role of the axis is inverted), but the value of $C_{0s}$ remains unchanged.\\
\indent The contact ellipticity $\rho_s$ is
\begin{equation}
	\rho_s=\frac{b}{a}=\sqrt{\frac{1-\psi_s}{1+\psi_s}}  \,.
	 \label{eq:rho_s}
\end{equation}
It is independent of the normal load and of the curvature radius of the indenter. The penetration depth, $\delta$, is obtained from the condition $H\left( 1 \right)=0$ in Eq.~(\ref{eq:H}):
\begin{align}
	\delta&=\frac{c^2}{R}\label{eq:delta}\\
	&=\frac{a^2}{R}\frac{1+\rho_s^2}{2} \label{eq:delta_a}\\
	&=\frac{b^2}{R}\frac{1+\rho_s^{-2}}{2}\label{eq:delta_b}
\end{align}
Then
\begin{equation}
 H\left( s \right)=\frac{3K}{8\lambda_x^2\lambda_y^2C_{0s}R}\frac{c^3}{ab}\left( 1-s^2\right) 
\end{equation} 
The normal load $P$ is obtained from (\ref{eq:PdeH}) as
\begin{equation}
 P=\frac{1}{\lambda_x^2\lambda_y^2 C_{0s}}\frac{c^3K}{ R} \label{eq:P}
\end{equation} 
It can also be expressed using the semi-axis lengths as
\begin{align}
	P&=\frac{ \left( 1+\rho_s^2 \right)^{\frac{3}{2}}}{2^{\frac{3}{2}}\lambda_x^2\lambda_y^2C_{0s}}\frac{a^3K}{R} \label{eq:P_a}\\
	&=\frac{ \left( 1+\rho_s^{-2} \right)^{\frac{3}{2}}}{2^{\frac{3}{2}}\lambda_x^2\lambda_y^2C_{0s}}\frac{b^3K}{R} \label{eq:P_b} 
\end{align}
and the load-penetration law reads
\begin{equation}
	P=\frac{KR^{\frac{1}{2}}\delta^{\frac{3}{2}}}{\lambda_x^2\lambda_y^2C_{0s}} \,.
\end{equation}
\indent The contact stiffness, $S$, which is known to be affected by initially stressed state~\cite{yang2004}, can be obtained by differentiating the previous expression with respect to the penetration depth. Expressing the result in terms of the contact size parameter $c$, one obtains
\begin{equation}
	S=\frac{3}{2}\frac{Kc}{\lambda_x^2\lambda_y^2C_{0s}}\,.\label{eq:stiff}
\end{equation}
Eqs.~(\ref{eq:P}), (\ref{eq:delta}) and (\ref{eq:stiff}) are similar to the corresponding ones for the Hertzian contact on an un-stretched substrate, provided that the contact radius  and the reduced modulus are respectively replaced with an averaged contact size, $c$, and by the quantity $K\left( \lambda_x^2\lambda_y^2C_{0s} \right)^{-1}$.\\
\indent From Eqs.~(\ref{eq:P_a}) and (\ref{eq:P_b}), at a given stretch state, both semi-axis lengths are observed to follow a Hertzian-like dependence on the normal load, but with different effective moduli. This result could have been anticipated using dimensional analysis (see Appendix~\ref{app:dim_analysis}).\\
\indent When the substrate is un-stretched, one obtains a circular contact area: $\varphi=0$ in (\ref{eq:eq}) leads to $\psi_s=0$. Thus $\rho_s=1$ and, as $\bar{\eta}=1$, one obtains $ C_{0s}=1$. The classical results for Hertzian contacts are retrieved from expressions (\ref{eq:P_a}), (\ref{eq:P_b}), (\ref{eq:delta_a}) and (\ref{eq:delta_b}). For equiaxially stretched substrate, the contact remains circular too. In this case, however, as $\bar{\eta}=\lambda^3$, if $\lambda$ is the stretch and $a$ the contact radius,
\begin{align}
	P&=\frac{a^3K'\left( \lambda \right)}{R}\\
	\delta&=\frac{a^2}{R} \,,
\end{align}
where
\begin{equation}
	K'\left( \lambda \right)=\frac{\lambda^9+\lambda^6+3\lambda^3-1}{2\lambda^4\left( \lambda^3+1 \right)}K \,.
\end{equation}
\indent Thus, in the case of equiaxial stretch, the contact is Hertzian, with a stretch-dependent effective modulus. It can be verified that it is higher (resp. lower) than the substrate elastic modulus when there is traction (resp. compression) in the contact plane. These results are in agreement with the analysis of adhesive contact of a sphere on a equiaxially stretched substrate~\cite{he2009} when adhesion is neglected.
\subsection*{Adhesive contact}
When $H\left( 1 \right)\ne 0$, normal stress can be derived from Eqs.~(\ref{eq:H}) and (\ref{eq:contr}) as
\begin{equation}
	p\left( u \right)=\frac{2}{\pi}\frac{H\left( 1 \right)}{\sqrt{1-u^2}}+\frac{3K}{4\pi RC_2\lambda_x^2\lambda_y^2}\frac{c^3}{ab}\sqrt{1-u^2} \,.
	\label{eq:sigadh}
\end{equation}
\indent A similar form was postulated in the case of adhesive contact of a sphere on an equiaxially stretched substrate by He and Dong~\cite{he2009}. In their analysis, the parameter $c$ would represent the actual radius of the circular contact area. In line with the JKR description of adhesion, the above expression represents the superposition of a rigid cylindrical punch displacement and of a non-adhesive indentation. The constant $H\left( 1 \right)$ may be determined from an appropriate estimation of the adhesion energy~\cite{johnson1985,he2009}. Alternatively, Maugis and Barquins \cite{maugis1981,barquins1982} used a fracture mechanics argument jointly with Griffith criterion to fix the value of the constant. In the following, a method equivalent to the former one is used.\\
\indent It should be noticed first that the isotropy condition for the displacements in the contact area is the same as for the non-adhesive case. Indeed, the second term in Eq.~(\ref{eq:sigadh}) induces normal displacements with a square dependence with the radius while the first one only contributes to a rigid displacement of the points within the contact area. The later is equivalent to the stress field associated with a cylindrical punch and does not add any in-plane anisotropy. Then, as in the non-adhesive case, the value $\psi_s$ for the anisotropy parameter $\psi$ insures the isotropy of the displacements. It follows that adhesion has no effect on the eccentricity of the contact area.\\
\indent From eq. (\ref{eq:H}), the constant $H(1)$ is related to the indentation depth by
\begin{equation}
	H\left( 1 \right)=\frac{3K}{8C_{0s}\lambda_x^2\lambda_y^2}\frac{c}{ab}\left( \delta-\frac{c^2}{R} \right) \,.
\end{equation}
Stress function can be written as
\begin{equation}
	p\left( u \right)=\frac{3K}{4\pi C_{0s}\lambda_x^2\lambda_y^2}\frac{c}{ab}\left[ \frac{ \delta-\frac{c^2}{R}}{\sqrt{1-u^2}}+\frac{2c^2}{R}\sqrt{1-u^2}\right] \,.
\end{equation} 
Indentation induced stored elastic energy is
\begin{align}
	U_{el}&=\frac{1}{2}\iint \left( \delta-\frac{x^2+y^2}{2R} \right)p\left( \sqrt{\frac{x^2}{a^2}+\frac{y^2}{b^2}}\right)\,dxdy\\
	&=\frac{ab}{2}\int_0^{2\pi}d\theta\int_0^1\left( \delta-s^2\frac{a^2\cos^2\theta+b^2\sin^2\theta}{2R} \right)p\left( s\right)s\,ds\\
	&=\frac{K c}{20 R^2C_{0s}\lambda_x^2\lambda_y^2}\left( 15R^2\delta^2-10Rc^2\delta+3c^4 \right) \,.
\end{align}
Under equilibrium conditions, the relation $\left( \frac{\partial U_{el}}{\partial A} \right)_\delta=w$ holds, with $A=\pi ab$. As anisotropy of the contact shape is kept constant,
\begin{equation}
	\left( \frac{\partial U_{el}}{\partial A} \right)_\delta=\frac{\sqrt{1+\rho_s^2}}{2\sqrt{2}\pi\rho_sa}\left( \frac{\partial U_{el}}{\partial c} \right)_\delta
\end{equation} 
and thus
\begin{equation}
	w=\frac{\sqrt{1+\rho_s^2}}{\rho_sa}\frac{3\sqrt{2} K }{16 \pi C_{0s}\lambda_x^2\lambda_y^2}\left( \delta-\frac{c^2}{R} \right)^2 \,.
\end{equation}
Integration of the normal stress gives
\begin{equation}
	P=\frac{K c}{2C_{0s}\lambda_x^2\lambda_y^2}\left(3 \delta-\frac{c^2}{R} \right) \,.
\end{equation}
Defining $P_1$, the load corresponding to the same contact area in the non-adhesive case (\ref{eq:P}) as 
\begin{equation}
	P_1=\frac{K c^3}{C_{0s}\lambda_x^2\lambda_y^2R}=\frac{ \left( 1+\rho_s^2 \right)^{\frac{3}{2}}}{2^{\frac{3}{2}}\lambda_x^2\lambda_y^2C_{0s}}\frac{a^3K}{R} \,,
\end{equation}
one obtains
\begin{equation}
	\frac{\left( P_1-P \right)^2}{6\pi a^3 K}=\frac{ w\rho_s}{C_{0s}\lambda_x^2\lambda_y^2}\sqrt{\frac{1+\rho_s^2}{2}} \,.
\end{equation}
The form of this relation is equivalent to the JKR expression and it constitutes its generalization to te case of an incompressible stretched substrate. Here again, the unstretched situation corresponding to JKR model is retrieved by letting $C_{0s}=1, \rho_s=1$.\\ 
\indent For equiaxially stretched substrate, the previous expression reads
\begin{equation}
	\frac{\left( P_1-P \right)^2}{6\pi a^3 K'\left( \lambda \right)}=w \,,
\end{equation}
where
\begin{equation}
	P_1=\frac{a^3K'\left( \lambda \right)}{R}
\end{equation}
and
\begin{equation}
	K'\left( \lambda \right)=\frac{\lambda^9+\lambda^6+3\lambda^3-1}{2\lambda^4\left( \lambda^3+1 \right)}K \,.
\end{equation}
The usual JKR relation is obtained, with an effective modulus which depends on the stretch ratio. An equivalent expression was obtained in ref.~\cite{he2009}.
\section*{Experimental results}
\label{sec:discussion}
\begin{figure}[ht]
	\begin{center}
		\includegraphics[width=0.9\columnwidth]{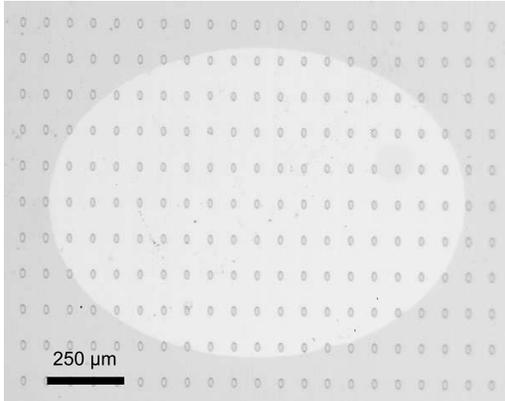}
		\caption{Elliptical contact shape recorded for a stretch ratio $\lambda$=1.32 and a contact load $P=$100~mN (the stretching is applied along the vertical direction).} 
		\label{fig:contact}
	\end{center}
\end{figure}

\begin{figure}[h!]
	\centering
	\includegraphics[width=0.9\columnwidth]{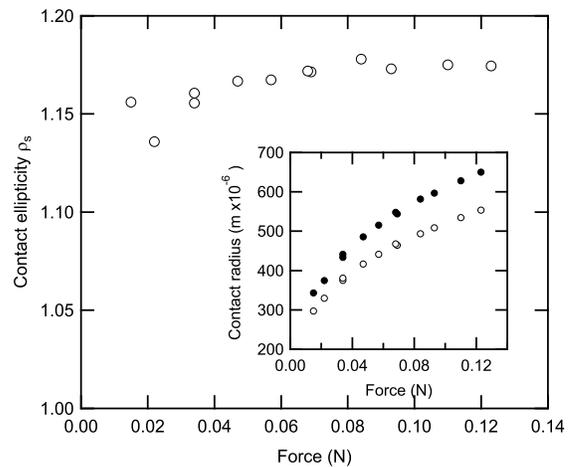}
	\caption{Ellipticity ratio $\rho_s=b/a$ of the contact area as a function of the applied contact load for a stretch ratio $\lambda=1.191$. Inset: corresponding values of ($\bullet$) the major and ($\circ$) the minor semi-axis lengths of the elliptical contact.}
	\label{fig:rho_fn}%
\end{figure}

\begin{figure}
	\centering
	\includegraphics[width=0.9\columnwidth]{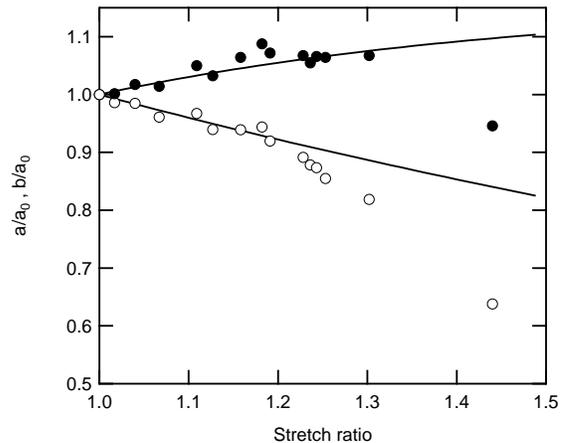}
	\caption{Semi-axis lengths ($\circ$) $a$ and $b$ ($\bullet$) of the elliptical contact as a function of the stretch ratio $\lambda$. Solid lines correspond to the theoretical prediction of Eqn.~(\ref{eq:P_a}) and (\ref{eq:P_b}).}
	\label{fig:a_b_lambda}%
\end{figure}

\begin{figure}
	\centering
	\includegraphics[width=0.9\columnwidth]{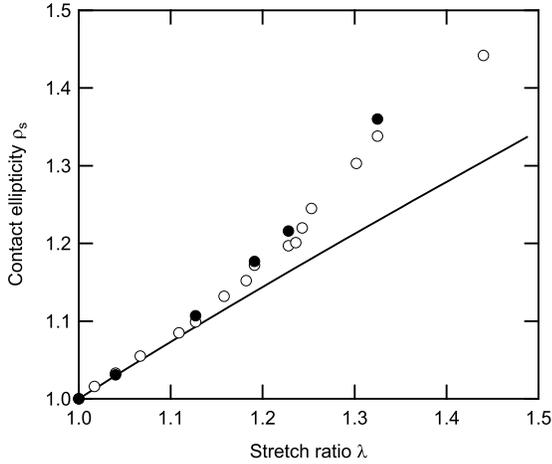}
	\caption{Ellipticity ratio $\rho_s=b/a$ of the contact area as a function of the stretch ratio $\lambda$. ($\circ$) contact in air; ($\bullet$) contact in water. Solid line: theoretical prediction of Eq.~(\ref{eq:eq}).}
	\label{fig:rho_lambda}%
\end{figure}

\begin{figure}[h]
	\centering
	\includegraphics[width=0.9\columnwidth]{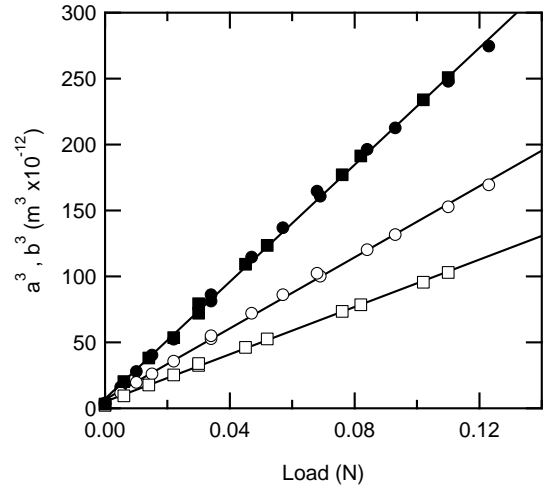}
	\caption{Cube of the semi-axis lengths $a$ (open symbols) and $b$ (filled symbols) of the elliptical contact as a function of the applied load. Circles: $\lambda=1.19$, squares: $\lambda=1.32$. Solid lines correspond to linear regression fits.}
	\label{fig:a3_b3_fn}%
\end{figure}

An example of a contact area picture is shown in Fig.~\ref{fig:contact} for a stretch ratio $\lambda$=1.32 and a normal load $P=$100~mN. 
The elliptical shape of the contact is clearly evidenced with the major axis of the ellipse perpendicular to the stretch direction. Consistently with a numerical analysis of Eq.~(\ref{eq:rho_s}) with $\lambda_x=\lambda,\lambda_y=\lambda^{-\frac{1}{2}}$, the aspect ratio is larger than one ($b>a$). Fig.~\ref{fig:rho_fn} shows the changes in the contact ellipticity $\rho_s=b/a$ as a function of the applied contact load for a stretch ratio $\lambda=\lambda_x=1.19$. consistently with the prediction for a non adhesive contact (Eqs~(\ref{eq:eq}) and (\ref{eq:rho_s})), it turns out that the measured aspect ratio is nearly constant over the whole investigated load range. This feature was preserved for all the stretch ratios under consideration.\\
\indent The dependence of the contact shape on the stretch ratio is further examined in Figures~(\ref{fig:a_b_lambda}) and (\ref{fig:rho_lambda}) where $a$,$b$ and their ratio $\rho_s$ are reported as a function of $\lambda$. In these figures, the black solid lines correspond to theoretical predictions. For stretch ratios less than about 1.2, experimental data are found to be in good accordance with model. Above this threshold, deviations from the theoretical predictions can be attributed to departure of the PDMS mechanical behavior from the Neo-Hookean description which is embedded in the model (cf~Fig.~(\ref{fig:stress_stretch})).\\
\indent The effects of adhesion on the stretch dependence of the contact ellipticity were considered by carrying out some of the experiments with the contact fully immersed in a droplet of deionized water. As indicated in Appendix~A, this resulted in a decrease in the adhesion energy of the unstretched substrate from 27~mJ~m$^{-2}$ to 5~mJ~m$^{-2}$. A comparison between experiments carried out both in air and in water (Fig.~(\ref{fig:rho_lambda})) show that the $\rho_s(\lambda)$ relationship is insensitive to such a change in adhesion. As stressed in the theoretical section, the isotropy condition for the displacements in the contact area is enforced by both adhesive and non-adhesive contacts. As a consequence adhesion does not add any in plane anisotropy and the ellipticity remains load-independent, consistently with experimental observations.\\
\indent The load dependence of the semi-axis lengths was further examined in the light of Eqs.~(\ref{eq:P_a}) and (\ref{eq:P_b}) which predict a linear dependence of $a^3$ and $b^3$ on $F$. In Fig.~(\ref{fig:a3_b3_fn}), the measured values of $a^3$ and $b^3$ are reported as a function of applied load for two values of the stretch ratio ($\lambda=1.19$ and $\lambda=1.32$). A linear behavior of both $a^3(F)$ and $b^3(F)$ relationships is indeed observed. The vanishing intercept of the linear fits indicates that adhesive effects can be neglected in such a representation. It can also be noted that the semi-axis length $a$ for a given load is only marginally affected by the stretch ratio difference while a more pronounced effect is observed for the semi-axis length $b$.\\
\indent According to Eqns.~(\ref{eq:P_a}) and (\ref{eq:P_b}), the slopes of these linear relationships provide an estimate of the 'effective' moduli $K_{eff}^a$ and $K_{eff}^b$ along the major and minor axis of the contact ellipse which can be defined as follows
\begin{align}
	K_{eff}^a&=K \frac{ \left( 1+\rho_s^2 \right)^{\frac{3}{2}}}{2^{\frac{3}{2}}\lambda_x^2\lambda_y^2C_{0s}} \label{eq:keffa} \,,\\
	K_{eff}^b&=K \frac{ \left( 1+\rho_s^{-2} \right)^{\frac{3}{2}}}{2^{\frac{3}{2}}\lambda_x^2\lambda_y^2C_{0s}} \label{eq:keffb}.
\end{align}
For $\lambda=1.19$, linear fits to data provides $K_{eff}^a/K=1.29 $ and $K_{eff}^a/K=0.79 $, in relatively good agreement with Eqns.~(\ref{eq:P_a}) and (\ref{eq:P_b}) which predict $K_{eff}^a/K=1.20$ and $K_{eff}^a/K=0.86 $. Conversely, the deviation from the neo-Hokeean behaviour results in a significant departure from the theory for $\lambda=1.325$, especially regarding $K_{eff}^a$ value which is under-predicted by about 25\%. 
\section*{Concluding remarks}
The present model was derived for a spherical probe and it should be mentioned that it cannot directly be extended to axisymmetrical punch shapes other than spherical or cylindrical. As an example, for a conical shape with semi-angle $\alpha$, assuming a stress function similar to (\ref{eq:pression}) imply to solve an integral equation which is equivalent to Eq. (\ref{eq:eqint}):
\begin{equation}
	\delta-\frac{r}{\tan \alpha}=\frac{8\lambda_x^2\lambda_y^2}{3\pi K}\frac{ab}{c}\int_0^{2\pi}\frac{\bar{\eta}+1}{\bar{\omega}\bar{\gamma}}H\left(\frac{r}{c}\frac{ \left|\cos\left( \beta-\theta \right)\right|}{\bar{\gamma} }\right)\,d\beta
\end{equation} 
The solution $H(s)$ is linear and can be accepted if the coefficient
\begin{equation}
 C_1=\frac{1}{\pi}\int_0^{2\pi}\frac{\bar{\eta}+1}{\bar{\omega}\bar{\gamma}^2}\left|\cos\left( \beta-\theta \right)\right|\,d\beta  \label{eq:C1}
\end{equation} 
can be made independent of $\theta$ by a proper choice of the parameter $\psi$. This is clearly the case when the substrate is isotropically stretched ($\lambda_x=\lambda_y$, $\phi=0$): in this case $\psi=0$  ($a=b$) is a solution. However, in the general case where $\lambda_x\ne\lambda_y$, considering that the Fourier series of
\begin{equation}
 \left|\cos\left( \beta-\theta \right)\right|=\sum_{m=0}^\infty A_m\cos\left( 2m\left( \beta-\theta \right) \right)
\end{equation} 
and noting that $\left( \bar{\eta}+1 \right)\bar{\omega}^{-1}\bar{\gamma}^{-2}$ is a function of $\cos 2\beta$, the coefficient can be written as
\begin{equation}
 C_1=\sum_{m=0}^\infty B_m \cos 2m\theta \label{eq:sfC1} \,,
\end{equation}
where
\begin{equation}
 B_m=\frac{A_m}{\pi}\int_0^{2\pi}\frac{\bar{\eta}+1}{\bar{\omega}\bar{\gamma}^2}\cos\left( 2m\beta\right)\,d\beta \,. 
\end{equation}
\indent The invariance with respect to $\theta$ of the coefficient would require that the Fourier series (\ref{eq:sfC1}) is constant, or, equivalently, $B_m=0$ for $m=1,2,3...$. In turn this would require $\left( \bar{\eta}+1 \right)\bar{\omega}^{-1}\bar{\gamma}^{-2}$ to be constant, which is impossible when $\varphi\ne 0$ (or $\lambda_x\ne\lambda_y$). This argument also applies for punch profiles $\sim r^p$ when $p$ is an odd integer. When $p$ is even, a finite number of harmonics contribute. However $p=2$ (the spherical shape) is the only case where the cancellation of a single harmonics of the function is sufficient, leading to the exact solution presented in this paper.\\
\indent Though exact solution cannot be found for an arbitrary punch profile, approximate solutions may be derived considering that the difference between actual contact and an elliptic shape is not expected to be large. In the case of the conical indenter, for instance, we enforce that the values of the coefficient $C_1$ (Eq. (\ref{eq:C1})) should be identical along the directions $x$ and $y$. Alternatively, we can enforce that the leading anisotropic term of the Fourier expansion (\ref{eq:sfC1}), $B_1$, is zero:
\begin{equation}
 \int_0^{2\pi}\frac{\bar{\eta}+1}{\bar{\omega}\bar{\gamma}^2}\cos\left( 2\beta\right)\,d\beta =0
\end{equation}
\indent For an uniaxially stretched substrate, numerical calculations  shows that both approximations are in very good agreement and that $a/b\simeq \lambda$ within few percents when $\lambda< 1.5$. Once the ellipticity parameter is determined, expressions relating normal load and displacement or contact size are obtained in a way similar to the case of the spherical indenter. To be more specific, above discussion was dealing with the conical punch but it can be extended to any axisymmetric indenter with a power law profile.\\
\indent Some additional comments are also in order regarding elastic contact theories dealing with general Hertzian elliptical contacts. In this context, Johnson and Greenwood developed an approximate theory for adhesive elliptical contacts by expressing that the stress intensity factor remains almost constant all around the contact periphery \cite{johnson2005}. They concluded that eccentricity varies with the load. In the case of an adhesive stretched substrate, above results indicate that the contact area for a sphere is  elliptical, but its eccentricity is load-independent and identical to the Hertzian case. The stress intensity factor varies along the contact edge, reflecting  the anisotropy of the pre-stretched substrate properties upon incremental displacements.\\
\indent In conclusion, we have shown that contact area of a un-deformable sphere on a stretched elastomeric substrate has an elliptic shape. Its eccentricity is completely determined by the in-plane stretch of the substrate. In particular, it neither depends on the applied load, nor on the curvature radius of the sphere or on adhesion properties. 
\begin{acknowledgments}
 The authors wish to thank L. Olanier for his help in the design and in the realization of the contact device.\\
\end{acknowledgments}
%
\appendix

\section{Dimensional analysis of the stretched contact}
\label{app:dim_analysis}
According to Eqs.~(\ref{eq:P_a}) and (\ref{eq:P_a}), both semi-axis lengths are observed to follow a Hertzian-like dependence on the normal load at a given stretch state, but with different effective moduli. This result could have been anticipated using dimensional analysis. It can be observed that the semi-axis length, $a$, is determined when the modulus, $K$, the normal load, $P$, the sphere radius, $R$, and the stretch state, $\lambda_x,\lambda_y$  are fixed. Forming non-dimensional numbers, these parameters must be linked as 
\begin{equation}
	\frac{P}{Ka^2}=f_a\left( \frac{a}{R},\lambda_x,\lambda_y \right)
\end{equation} 
Now, keeping the contact area constant, if all normal displacements are multiplied by a number $\chi$, by linearity of the stress-strain problem described by the Green tensor, the normal load is also multiplied by $\chi$. From simple geometrical considerations, it correspond to a contact problem for a sphere with a radius divided by $\chi$. Thus
\begin{equation}
	\chi\frac{P}{Ka^2}=f_a\left( \chi\frac{a}{R},\lambda_x,\lambda_y \right)
\end{equation} 
which should hold for any values of $\chi$. The function $f_a$ is thus homogeneous and of the first degree with respect to $a/R$. One may deduce
\begin{equation}
	P=\frac{a^3K}{R}g_a\left( \lambda_x,\lambda_y \right) \,.
\end{equation} 
Similar results can be obtained for the other axis and for the penetration depth. One have also
\begin{equation}
	\frac{b}{a}=\frac{g\left( \lambda_x,\lambda_y \right)}{g\left( \lambda_y,\lambda_x \right)} \,.
\end{equation} 
which expresses that the contact shape anisotropy is controlled by the stretch anisotropy only.
It can be noticed that, in the un-stretched situation, previous considerations, allow to recover  the Hertzian expressions up to a constant numerical factor.

\section{Adhesive contact of the unstretched substrate}
\label{app:adhesive_contact}
The adhesive contact between the unstretched ($\lambda=1$) PDMS substrate and the glass lens was examined within the framework of the JKR theory~\cite{johnson1971}. The following linearised form of the relationship between the contact radius $a$ and the applied load $F$ was considered
\begin{equation}
 \frac{a^{3/2}}{R}=\frac{1}{K}\frac{F}{a^{3/2}}+\sqrt{\frac{6 \pi w}{K}} \,.
\end{equation}

\begin{figure}
 	\centering
	\includegraphics[width=0.9\columnwidth]{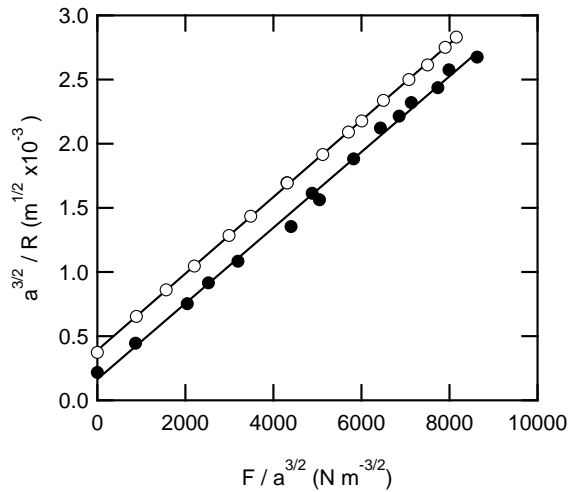}
	\caption{Linearised plot giving the contact radius $a$ versus the applied normal force $F$ for an unstretched ($\lambda=1$) PDMS substrate. ($\circ$) contact in air, ($\bullet$) contact immersed in deionized water. Solid lines are linear regression fits.}
	\label{fig:JKR}%
\end{figure}

\indent The slope of this linearised relationship thus provides the value of the reduced modulus $K$ while the adhesive energy $w$ is deduced from the intercept. In Fig.~(\ref{fig:JKR}), experimental contact results have been reported using this representation for contacts either in air or fully immersed in a droplet of deionized water. As expected, linear relationships are obtained in both case with the same slope, i.e. the same value of the reduced modulus ($K=3.36 \pm 0.02$~MPa). On the other hand, a decrease in the intercept is observed for contact in air which reflects a decrease in the adhesion energy from $w=27$~mJ/m$^2$ (air) to $w=5$~mJ/m$^2$ (water) as a result of the screening of van der Walls forces between surfaces by the water molecules.
%
%
\bibliographystyle{unsrt}
%

\end{document}